%% file: main.tex
\documentclass[lettersize,journal]{IEEEtran}
\usepackage{amsmath,amsfonts}
\usepackage{algorithmic}
\usepackage{array}
\usepackage[caption=false,font=normalsize,labelfont=sf,textfont=sf]{subfig}
\usepackage{textcomp}
\usepackage{stfloats}
\usepackage{url}
\usepackage{xcolor}
\usepackage{verbatim}
\usepackage{graphicx}
\usepackage{booktabs}
\usepackage{float}
\usepackage{url}
\usepackage{amssymb}
\usepackage{pifont}
\usepackage{multirow}
\usepackage{adjustbox}
\newcommand{\cmark}{\ding{51}}%
\newcommand{\xmark}{\ding{55}}%
\def\BibTeX{{\rm B\kern-.05em{\sc i\kern-.025em b}\kern-.08em
    T\kern-.1667em\lower.7ex\hbox{E}\kern-.125emX}}
\usepackage{balance}

\input{macro}

\begin{document}
\title{MT2KD: Towards A General-Purpose  Encoder for Speech, Speaker, and Audio Events}


\author{Xiaoyu Yang,~\IEEEmembership{Student Member,~IEEE,}, Qiujia Li,~\IEEEmembership{Member,~IEEE,}\\Chao Zhang,~\IEEEmembership{Member,~IEEE,} Philip C. Woodland,~\IEEEmembership{Fellow,~IEEE}
\thanks{Xiaoyu Yang and Philip C. Woodland are with the Department of Engineering, University of Cambridge, Trumpington St., CB2 1TQ Cambridge, U.K.
(e-mail: xy316@cam.ac.uk; pcw@eng.cam.ac.uk). Chao Zhang is with the Department of Electronic Engineering, Tsinghua
University, Beijing 100190, China (e-mail: cz277@tsinghua.edu.cn). Qiujia Li is with Google DeepMind (e-mail: qiujia@google.com). \\
This work has been submitted to the IEEE for possible publication. Copyright may be transferred without notice, after which this version may no longer be accessible.}
}


\maketitle

\input{0.Abstract}


\input{1.Intro}
\input{2.Related}
\input{3.Method}
\input{4.Setup}

\input{5.Results}

\input{6.Conclusions}

\bibliographystyle{IEEEtran}
\bibliography{ref.bib}

\end{document}

%% file: macro.tex
\newcommand{\ignore}[1]{}

\makeatletter
\DeclareRobustCommand\onedot{\futurelet\@let@token\@onedot}
\def\@onedot{\ifx\@let@token.\else.\null\fi\xspace}

\makeatother

\definecolor{MyDarkBlue}{rgb}{0,0.08,1}
\definecolor{MyDarkGreen}{rgb}{0.02,0.6,0.02}
\definecolor{MyDarkRed}{rgb}{0.8,0.02,0.02}
\definecolor{MyDarkOrange}{rgb}{0.40,0.2,0.02}
\definecolor{MyPurple}{RGB}{111,0,255}
\definecolor{MyRed}{rgb}{1.0,0.0,0.0}
\definecolor{MyGold}{rgb}{0.75,0.6,0.12}
\definecolor{MyDarkgray}{rgb}{0.66, 0.66, 0.66}

\usepackage{pgfplots}
\DeclareUnicodeCharacter{2212}{−}
\usepgfplotslibrary{groupplots,dateplot}
\usetikzlibrary{patterns,shapes.arrows}
\pgfplotsset{compat=newest}

%% file: 0.Abstract.tex
\begin{abstract}
    With the advances in deep learning, the performance of end-to-end (E2E) single-task models for speech and audio processing has been constantly improving.
    However, it is still challenging to build a general-purpose model with high performance on multiple tasks, since different speech and audio processing tasks usually require different training data, input features, or model architectures to achieve optimal performance. 
    In this work, MT2KD, a novel two-stage multi-task learning framework is proposed to build a general-purpose speech and audio encoder that jointly performs three fundamental tasks: automatic speech recognition (ASR), audio tagging (AT) and speaker verification (SV). In the first stage, multi-teacher knowledge distillation (KD) is applied to align the feature spaces of three single-task high-performance teacher encoders into a single student encoder using the same unlabelled data. In the second stage, multi-task supervised fine-tuning is carried out by initialising the model from the first stage and training on the separate labelled data of each single task. 
    Experiments demonstrate that the proposed multi-task training pipeline significantly outperforms a baseline model trained with multi-task learning from scratch. The final system achieves good performance on ASR, AT and SV: with less than 4\% relative word-error-rate increase on ASR, only 1.9 lower mean averaged precision on AT and 0.23\% absolute higher equal error rate on SV compared to the best-performing single-task encoders, using only a total of 66M model parameters. 

\end{abstract}

%% file: 1.Intro.tex
\section{Introduction}
\noindent End-to-end (E2E) trainable models have achieved excellent performance on a wide range of speech and audio processing tasks. The model architectures and training pipelines are constantly evolving, leading to models even surpassing human experts~\cite{spille2018comparing}. 
Due to the distinct task purposes, different speech and audio processing tasks tend to require quite different model architectures and training data to achieve the best performance. 

Although high performance on each task can be achieved using separate single-task models, it is still desirable to construct a single model performing multiple speech and audio processing tasks. Such a model could have the following advantages. First, a multi-task model is both computation-efficient and parameter-efficient since the computation and parameters are shared by multiple tasks. Second, multi-task learning may create task synergy, leading to improved performance over single-task models. Third, with the breakthrough in the development of artificial general intelligence \cite{bubeck2023sparks}, general-purpose audio models are of increasing interest due to their ability to unify multiple tasks \cite{hubert, wavlm, salmonn}.
However, multi-task learning of different speech and audio processing tasks faces the following issues. First, the model architecture used is different across different tasks. Task-specific architectures are employed in single-task models to achieve optimal performance and using a single model architecture could lead to sub-optimal performance on each task. Second, speech processing and audio processing tasks require different training data, and coordinating multi-task learning is challenging. Third, the objective of different tasks could interfere with each other, leading to reduced task synergy. For example, an automatic speech recognition (ASR) model should output the same text regardless of the speaker identity, whereas speaker verification (SV) aims at distinguishing different speakers regardless of the text spoken. Recently, self-supervised learning (SSL) leveraging vast amounts of unlabelled speech has achieved promising results on various speech processing tasks using the same pre-trained backbone encoder~\cite{wavlm, superb}, indicating that the feature space of various speech processing tasks can be aligned. Whisper~\cite{whisper} performs multi-task training of two speech processing tasks: ASR and automatic speech translation. However, other audio processing tasks are not included in its training pipeline.

Knowledge distillation (KD) is a common approach to improve the performance or parameter efficiency of single-task models, where the student model is trained to match the output of a teacher model instead of using labelled data. The teacher model is usually a powerful model trained via self-supervised learning~\cite{1bestKD, ensembleKD_lhy} or supervised data~\cite{whisperKD} trained on a very large amount of data. For example, the one-best alignment of a fine-tuned Wav2vec 2.0 model~\cite{w2v2} can be extracted and the distribution along it was used as distillation target for a student model~\cite{1bestKD}. Distil-Whisper~\cite{whisperKD} used the pseudo labels generated by Whisper as supervision and obtains a smaller model with good performance. The output logits of a transformer audio tagging (AT) model were used to supervise a CNN model to improve its AT performance~\cite{AudioTaggingKD}.
Multiple teachers~\cite{ensembleKD_lhy, multi-teacher, CED} can be used to further improve the student model performance. In the field of ASR, ~\cite{multi-teacher} used embeddings extracted from 3 SSL pre-trained models as targets and minimized the L1 distance between the student embeddings and the targets. ~\cite{CED} used an ensemble of multiple audio tagging teacher models and improves the audio tagging accuracy of a student model.

Instead of naively performing multi-task learning from scratch using labelled data, we hypothesise that aligning all tasks in the feature space using unlabelled data could bridge the gap between different tasks more easily. In this work, we propose a multi-task multi-teacher KD framework to build a general-purpose audio encoder suitable for three fundamental speech and audio processing tasks: ASR, AT and SV. The training is split into two stages. In the first stage, KD training is carried out, where the student encoder model is supervised by three teacher models (one from each task) leveraging unlabelled data so that the feature space of the three tasks are aligned. In the second stage, the pre-trained student model is fine-tuned with supervised data to perform all three tasks. In particular, it is found that keeping the KD loss as an auxiliary loss leads to better fine-tuning results.

The main contributions of this paper are as follows:
\begin{enumerate}
    \item Proposed a two-stage multi-task multi-teacher KD training pipeline jointly performing ASR, AT and SV;
    \item Demonstrated that the multi-teacher KD pre-training using unlabelled data is necessary to align different tasks and leads to better performance on each task; 
    \item Conducted extensive experiments to analyse the relationship between ASR, AT and SV. In particular, we found that ASR and SV should be performed at different encoder depths to achieve a balance between the two tasks. 
\end{enumerate}

The remainder of this paper is organised as follows. Section~\ref{sec: related} gives a review of related work. Section~\ref{sec: method} introduces the details of the proposed general-purpose audio encoder. Section~\ref{sec: setup} and~\ref{sec: results} present the experimental setup and results. Section~\ref{sec: conclusions} draws conclusions.

%% file: 2.Related.tex
\section{Related Works}
\label{sec: related}

\noindent In this section, a brief introduction is given on three speech and audio processing tasks: ASR, AT and SV. The existing multi-task learning frameworks for different speech and audio processing tasks are reviewed. In addition, the applications of KD on the three tasks are introduced.

\subsection{Automatic Speech Recognition}

\noindent A typical E2E trainable ASR system consists of two components: an encoder that encodes the speech input into higher level feature representations and a decoder generating textual tokens. Filterbank features extracted at a certain frame-rate are commonly used as input features to the audio encoder.
Connectionist Temporal Classification~\cite{gravesCTC} was the first E2E trainable ASR architecture, which contains an audio encoder and a frame independent decoder (usually a linear layer). Later, the neural transducer~\cite{gravesRNNT} was proposed to address the frame-independence assumption in the CTC model. It consists of an audio encoder, a prediction network and a joint network. Given the frame-level input audio features $\mathbf{X}=\mathbf{x}_{1:T}$, the encoder generates contextualised audio representations. The predictor encodes the text sequence and the joint network fuses the output of both encoder and predictor and generates a probability distribution lattice. The training loss for a neural transducer is as follows:
\begin{align}
    \mathcal{L}_{\text{RNN-T}} = \sum_{a\in\mathcal{B}^{-1}(\textbf{y})} p(a|\mathbf{X}),
\end{align}
where $\textbf{y}$ is the target text sequence and $\mathcal{B}^{-1}(\textbf{y})$ is the set of all alignments mapping $\mathbf{X}$ to $\textbf{y}$.

Recently, ASR models trained on large-scale supervised data are showing superior performance. Whisper~\cite{whisper} utilises a 680K hours non-public weakly supervised training set covering a broad distribution of audio from many different environments, recording setups, speakers, and languages. The model shows very good performance on a wide range of ASR benchmarks. Meanwhile, the model also demonstrates stronger robustness to additive noise~\cite{whisper-at} compared to other models trained either by self-supervised learning or limited supervised data. 

\subsection{Audio Tagging}

\noindent In AT, a model receives an audio clip and predicts the type of sound event. Since a single audio clip can be categorised into multiple different classes, AT is usually treated as a multi-class classification task. 
Due to the capability of capturing short term information, convolutional neural networks (CNNs) achieved good performance~\cite{PANNs} and became a popular backbone model for AT~\cite{PANNs, CnnAT, DeepCNNAT}.
Gong et al.~\cite{AST} adopted a transformer model pre-trained on image classification task as the initialisation and trained the model on audio tagging data. The proposed audio spectrogram transformer (AST) adapts quickly to AT and outperforms previous CNN-based networks. AST splits the input log-mel filterbank features into square patches spanning over both time and frequency domain. An embedding is extracted for each patch, which is fed as input to the transformer backbone model. 
BEATs~\cite{BEATs} applied self-supervised pre-training for AT by generating discrete audio tokens via random projection quantisation~\cite{Best-RQ}. The pre-training of the quantiser and the backbone model is performed iteratively using unlabelled data. The backbone model follows the architecture in AST~\cite{AST}. In~\cite{BEATs}, the log-Mel filterbank features are split into $N$ 16x16 patches with an overlap of 6 in both time and frequency domains. The patch embeddings are extracted from each patch, which are fed to the transformer encoder as a sequence of token embeddings. The final prediction is made by averaging the logits $\textbf{p}_{1:N}$ of each patch followed by a sigmoid activation. The training loss $\mathcal{L}_{AT}$ for AT is formulated as follows:
\begin{align}
    \mathcal{L}_{\text{AT}} = \text{BCE}(C, \sigma(\sum_{i=1}^{N} \textbf{q}_i))
\end{align}
where \text{BCE} is the binary cross-entropy loss function for multi-class classification, $\sigma$ is the elementwise Sigmoid function and $C$ is the multi-hot audio event label.

Unlike ASR models that typically operate on frame-level acoustic representations, state-of-the-art AT models tend to use patch-level information so that both frequency and temporal information are considered, which are crucial for distinguishing different sound events. Experiments in ~\cite{PANNs, CnnAT} also confirm that a 2-D spectrogram produces better results than a 1-D waveform as input for AT.

\subsection{Speaker Verification}

SV is the task of determining if some speech is from a particular speaker, which is usually accomplished by measuring the similarity between the speaker embeddings extracted from the segment with an enrolled voice print. As for ASR, filterbank features are commonly used as input to SV. CNN-based networks demonstrated good performance on SV~\cite{xvector, ecapa, garcia2020jhu} due to their ability of capturing local patterns. 
ECAPA-TDNN is one of the most popular architectures, which consists of a frame-level features extractor (encoder) and a segment-level statistical self-attentive pooling module to aggregate the speaker embedding at the segment level. 
The encoder extracts high-dimensional frame-level acoustic features given the input features and the pooling module adopts a channel-dependent attention mechanism and re-weights the frame-level acoustic features.
The final speaker embedding is obtained by projecting the mean and standard deviation aggregated from the weighted frame-level embeddings to a fixed-dimensional vector. During training, a classification layer is apppended after the speaker embedding for predicting the speaker id and the whole model is trained with the cross-entropy loss. When performing speaker verification, the classification layer is discarded and only the speaker embeddings are used. If the cosine similarity between two speaker embeddings exceeds a pre-defined threshold, the two speakers are considered identical.

\subsection{Multi-task Learning of Speech and Audio Processing}

\noindent 


Multi-task learning has been explored for various audio processing tasks and achieved promising results. A multi-task audio encoder trained on 22 audio tasks was used as the initialisation of the audio encoder in the Clap framework~\cite{clap} and performs well on downstream audio processing tasks after fine-tuning~\cite{elizalde2023natural} . 

There has been a significant amount of work on multi-task training for speech processing~\cite{pironkov2016multi}. Multi-task training of speaker recognition and speech recognition is accomplished by utilising the inter-task features extracted from the task-specific branch~\cite{speech_mtl}. Whisper~\cite{whisper} is trained on various speech-processing tasks, including multilingual ASR, speech translation (X $\rightarrow$ English), spoken language identification and voice activity detection. \cite{whisper} demonstrated that multi-task training facilitates the learning of English ASR as the model size and training data scale up.  

However, few attempts have been made to build a multi-task model for both audio and speech processing tasks since they require very different training data and different model architectures. Whisper-AT~\cite{whisper-at} discovered that the audio encoder of the supervised-trained Whisper~\cite{whisper} is highly correlated to non-speech sounds and integrated the AT task in the encoder of Whisper. The layer-wise embeddings of the Whisper encoder were combined with a relatively small network in order to perform audio tagging.  When fine-tuned from a Whisper model, the model achieved a mean average precision (mAP) of 45.6 on AudioSet, surpassing existing models using frame-level audio features as input. However, if the encoder parameters are frozen so that the ASR capability is unaffected, the mAP drops to 41.3. Nonetheless, it still indicates that audio and speech processing tasks can be accomplished simultaneously.


\subsection{Knowledge Distillation in Speech and Audio Processing}

KD~\cite{HintonKD}, also known as teacher-student learning, is a widely used technique for model compression and performance improvement. In KD, the student model is trained to match the output of the teacher model, and different KD loss functions can be used depending on the form of teacher output, also called as teacher labels. 
Depending on how and when the teacher labels are generated, KD can be divided into two categories: off-line KD and on-line KD. In off-line KD, the teacher labels are pre-computed and stored at the data preprocessing stage, and are only accessed during KD training. In on-line KD, the teacher labels are computed on-the-fly, requiring extra computation and GPU memory consumption compared to off-line KD since the teacher model need to be forwarded in each batch. However, it also reduces the storage cost, especially for large corpora. 
Furthermore, consistent data augmentation~\cite{ConsistentKD} is easier to implement in on-line KD as long as the teacher and student share the same input data. 


KD has been extensively used in audio and speech processing tasks. A lot of research has been made to improve the accuracy or latency of ASR models using KD. Takashima~\cite{takashima2018investigation} et al used output distributions from a teacher CTC model to improve a student CTC model. ~\cite{mvq} attempted to predict the compressed representation of the embeddings extracted from self-supervised pre-trained models as auxiliary loss during ASR training.
Distil-Whisper~\cite{whisperKD} distils a Whisper model by generating pseudo labels using an open-source dataset and obtained a much smaller Whisper model with competitive performance. 
The speaker embedding generated by a teacher SV model was used as training targets to supervise a student SV model via mean squared error and cosine distance~\cite{SV_KD}

In AT, KD has been found to yield better performance than training on the ground-truth multi-hot labels~\cite{CED}. Consistent Ensemble Distillation (CED)~\cite{CED} adopts consistent KD in an off-line manner and utilised an ensemble teacher model and achieved state-of-the-art mAP of 50.0 on AudioSet~\cite{audioset}. To reduce the cost of storing augmented input audio features and teacher labels, the random seed for data augmentation and the corresponding teacher top-K logits are stored.

The student model performance is highly dependent on the teacher model. To improve the teacher model quality, ensemble distillation~\cite{ensembleKD_lhy} has been adopted, where a two-layer student model learns the representations from different layers of different self-supervised pre-trained models. The student model can be fine-tuned for various downstream tasks. Similarly, Yang et al~\cite{multi-teacher} proposed to distil a student ASR encoder using three fine-tuned self-supervised pre-trained ASR models, which produces a stronger student model than using a single teacher, indicating that the knowledge from multiple strong teachers could be complementary.

%% file: 3.Method.tex
\section{General-purpose Audio Encoder}
\label{sec: method}

\subsection{Motivation}

\noindent Although single task models can achieve good performance, there is still a demand on a general-purpose audio encoder that performs multiple speech/audio processing tasks simultaneously. Such a model is parameter and computation efficient, and could be a suitable foundation model for building speech-based large language models~\cite{salmonn}.
Naive multi-task learning of multiple speech processing tasks faces the following issues:
\begin{enumerate}
    \item Different tasks can requires \textbf{different types of input}. For example, typical ASR or SV systems require frame-level input features, whereas state-of-the-art AT systems operates on patch-level input at much higher granularity.
    \item Different tasks may have \textbf{distinct}, or even contradictory objective functions. For example, the output of an ASR system should be speaker-independent, whereas an text-independent SV system should only capture the speaker information instead of the spoken content, which is contradictory to the goal of ASR.
\end{enumerate}
Due to the above issues, naive multi-task learning may yield sub-optimal performance on each task. In Whisper-AT~\cite{whisper-at}, the multi-task model for ASR and AT achieved a much lower mAP than the single-task AT model. The performance of English ASR of a single-task and a multi-task model is compared~\cite{whisper} and performance degradation is observed when the amount training data is not sufficiently large.


In this work, we aim to build a general-purpose audio encoder suitable for various speech and audio processing tasks with minimal performance degradation on individual tasks. To this end, we propose to use KD to facilitate multi-task training. In particular, we focus on three tasks: ASR, AT and SV, as they address the fundamental problems in speech and audio processing. For each task, a dedicated teacher model is chosen. 
Since KD does not pose restrictions on the model architecture and the input data format, arbitrary combinations of teacher and student models can be adopted, resolving the difference in architecture design for different tasks. By using KD, the performance of the student model on each task can be improved due to the high-quality supervision provided by the teacher models. 
Since directly optimising distinct objective functions of different tasks may be difficult, we hypothesise that it is easier to align different tasks at the feature space and propose a two-stage training pipeline. 
In the first stage, the student encoder is pre-trained on three tasks via KD using only unlabelled data. During this stage, the student model is trained to match the feature representations generated by the three teacher models. By discarding the ground truth label in this stage, the mismatch between different objective functions affects training to a lesser extent.
After pre-training, the student encoder model is used to initialise a multi-task model in the second stage, which will be then fine-tuned with task-specific labelled data. Finally, the fine-tuned student model is capable of performing ASR, AT and SV at the same time.

\subsection{Naive Multi-task Training}

\noindent Before introducing the 2-stage multi-task KD training approach, the simplest multi-task supervised training framework for ASR, AT and SV is described. 
The input to the multi-task model is frame level audio features $\mathbf{X}$, and $\mathbf{y}$, $C$, $s$ are the groundtruth transcription, audio event and speaker ID for ASR, AT and SV respectively. 
Given the input acoustic features $\mathbf{X}$, the encoder outputs representations $\textbf{R}^n$ at each layer, where $n$ is the layer index ranging from 1 to the total number of encoder layers $N$. Three task-specific modules can utilise the audio representations at different layers of the encoder to perform each task.
The loss for multi-task training is formulated as follows:
\begin{align}
    \mathcal{L}_{\text{MTL}} &= \lambda_{1} \mathcal{L}_{\text{ASR}}(\textbf{R}^{i_{\text{ASR}}}, Y) \\
    & \:\:\:    + \lambda_{2} \mathcal{L}_{\text{AT}}(\textbf{R}^{i_{\text{AT}}}, C) \\
     & \:\:\:   + \lambda_{3} \mathcal{L}_{\text{SV}}(\textbf{R}^{i_{\text{SV}}}, s),
\end{align}
where $\mathcal{L}_{\text{ASR}}$, $\mathcal{L}_{\text{AT}}$, $\mathcal{L}_{\text{AT}}$ are the loss functions, $i_{\text{ASR}}$, $i_{\text{AT}}$ and $i_{\text{SV}}$ are the input layer indexes for ASR, AT and SV, respectively. The scales of KD losses for ASR, AT and SV are controlled by $\lambda_1$, $\lambda_2$ and $\lambda_3$.

\subsection{Multi-teacher Multi-task KD Pre-training} 

\begin{figure}
    \centering
    \includegraphics[width=\linewidth]{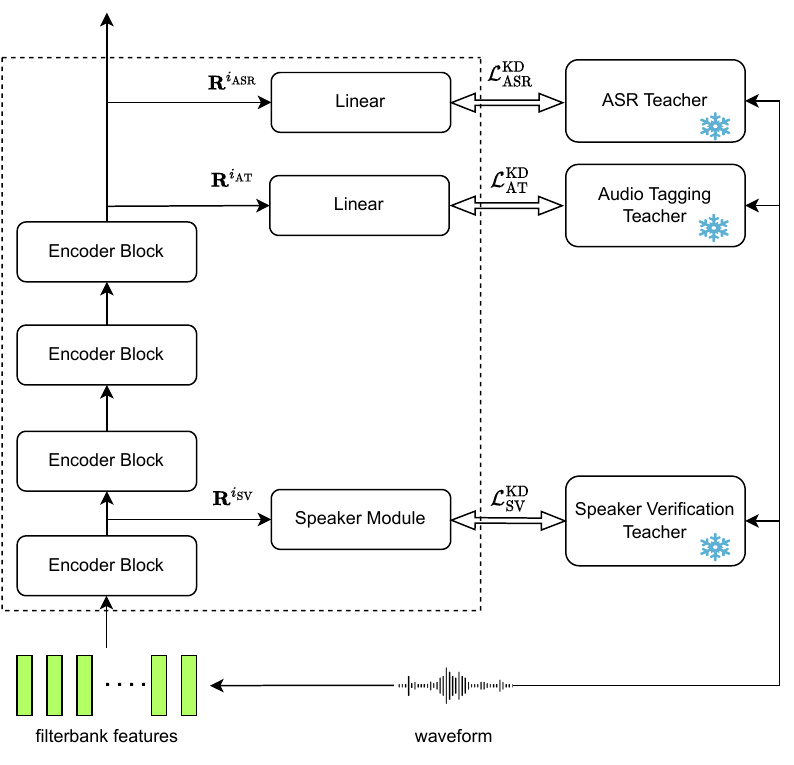}
    \caption{The multi-teacher multi-task KD pre-training framework. The feature extraction modules for each teacher are included in the teacher models. The teacher models are frozen all the time. The modules in the dashed box belong to the student model and the convolution subsampling module is emitted.}
    \label{fig:framework}
\end{figure}

\noindent Let $\mathcal{T}^{\text{ASR}}$, $\mathcal{T}^{\text{AT}}$ and $\mathcal{T}^{\text{SV}}$ be the teacher models for ASR, AT and SV respectively, we propose to use multi-teacher multi-task KD to improve the student encoder $\mathcal{S}$. Here, we assume $\mathcal{T}^{\text{ASR}}$ and $\mathcal{T}^{\text{SV}}$ operates on the frame-level input features and $\mathcal{T}^{\text{AT}}$ operates on the path-level input, as is the case for the state-of-the-art single task models. The student model is assumed to receive frame-level input features. An illustration of the KD-training framework is shown in Fig~\ref{fig:framework}.

Given the frame-level input features $\textbf{X}$, the student encoder generates frame-level contextualised speech representations $\textbf{R}^{i^{\mathcal{S}}}_{t}$ at each layer $i^{\mathcal{S}}$, where $t$ is the frame index. 
The encoder of $\mathcal{T}^{\text{ASR}}$ also produces frame-level speech representations $\textbf{TE}^{i^{\mathcal{T}}}$ at each layer $i^{\mathcal{T}}$. The frame-level L1 loss between the teacher speech representations at layer $i_{\mathcal{T}}$ and the student feature representations at $i_{\mathcal{S}}$ is formulated as:
\begin{align}
    \mathcal{L}^{\text{KD}}_{\text{ASR}} = \sum_{t=1}^{T} | \textbf{TE}^{i^\mathcal{T}_{\text{ASR}}}_t -\mathcal{M}_{\text{ASR}}(\textbf{R}^{i^\mathcal{S}_{\text{ASR}}}_t) |,
\end{align}
where $\mathcal{M}_{\text{ASR}}$ is a module transforming the student encoder representation to the same dimension of the teacher targets.
Since the teacher AT model operates on patch-level features, performing frame-level KD on AT using the teacher embeddings as the target as in ASR is infeasible. Due to the flexibility of KD, the K-dimensional output $C^{\mathcal{T}}$ of the logits before the final Sigmoid activation of $\mathcal{T}^{\text{AT}}$ can be used as teacher labels, where $K$ is the number of audio events. To perform AT, a dedicated module $\mathcal{M}_{\text{AT}}$ that takes the encoder outputs at $i_{\text{AT}}$-th layer also generates a C-dimensional logits. The binary cross-entropy loss between the teacher and student prediction is computed as follows:
\begin{align}
    \mathcal{L}^{\text{KD}}_{\text{AT}} = \text{BCE}(\sigma(C^{\mathcal{T}}), \sigma(\mathcal{M}_{\text{AT}}(\textbf{R}^{i_{\text{AT}}}))),
\end{align}
where $\sigma$ is the Sigmoid activation function and $i_{\text{AT}}$ the index of the input layer to $\mathcal{M}_{\text{AT}}$.

The output of $\mathcal{T}^{\text{SV}}$ is a $J$-dimensional speaker embedding. To perform KD on SV, a speaker embedding module $\mathcal{M}_{\text{SV}}$ taking the $i_{\text{SV}}$-th layer's output of $\mathcal{S}$ is appended, and the cosine similarity between the teacher and student speaker embeddings is minimized using the following loss:
\begin{align}
    \mathcal{L}^{\text{KD}}_{\text{SV}} = 1 - \textit{CosineSim}(\mathbf{v}, \mathcal{M}_{\text{SV}}(\textbf{R}^{i_{\text{SV}}})),
\end{align}
where $\mathbf{v}$ is the speaker embedding predicted by the teacher model and $\textit{CosineSim}$ is the cosine similarity between two un-normalized vectors. 

The final loss function for multi-task KD is formulated as below:
\begin{align}
    \mathcal{L}^{\text{KD}} = \lambda_1 \mathcal{L}^{\text{KD}}_{\text{ASR}} + \lambda_2 \mathcal{L}^{\text{KD}}_{\text{AT}} + \lambda_3 \mathcal{L}^{\text{KD}}_{\text{SV}},
\end{align}
where $\lambda_1$, $\lambda_2$ and $\lambda_3$ are the scaling factors for the loss functions of ASR, AT and SV, respectively. 
Despite that there is no dedicated multi-task dataset for ASR, SV and AT, computing three losses for one sample is still theoretically possible because ground truth labels are not needed during KD. However, only one loss based on the task from which the sample comes can also be computed.

\subsection{Multi-task Fine-tuning}

After the encoder is pre-trained with unlabelled data using the three teacher models, the model can be fine-tuned with task specific data to perform ASR, AT and SV. 
For ASR, the neural transducer architecture is adopted by appending a label predictor and a joint network after the audio encoder. 

For audio tagging, the weights of the linear layer projecting the encoder features to number of classes of audio events in the pre-training stage is kept as an initialisation. During fine-tuning, the binary cross-entropy loss is adopted as during pre-training.
For speaker verification, a linear layer projecting the speaker embeddings to number of speakers is used and the cross-entropy loss is adopted.

To avoid catastrophic forgetting during fine-tuning, the parameters of the pre-trained encoder are not updated during the initial warm-up period. Only the predictor, the joint network and the classification layer for SV are updated during this period. The whole model is then trained jointly, with the parameters of the audio encoder having a smaller learning rate for training stability. Note that during fine-tuning, the aforementioned KD loss functions can still be utilised as auxiliary losses to further improve the performance of the student model.

%% file: 4.Setup.tex
\section{Experimental Setup}
\label{sec: setup}

\subsection{Teacher Models}
\noindent The encoder of Whisper large-v3~\cite{whisper}, BEATs~\cite{BEATs} and ECAPA-TDNN~\cite{ecapa} are adopted as the teacher model for ASR, AT and SV, respectively. 
The encoder of the Whisper large-v3 model\footnote{\url{https://github.com/openai/whisper}} is a transformer model consisting of 24 transformer layers. It receives 128-channel log-magnitude mel spectrogram representation computed on 25~ms windows with a stride of 10~ms and generates 1280-D speech representations at a frame-rate of approximately 50~Hz. The Whisper model was trained on over 680,000 hours of weakly supervised data to perform ASR and automatic speech translation. Since Whisper is trained on a massive amount of supervised speech data under diverse acoustic conditions, its encoder representations could be beneficial for other tasks.
The BEATs models is an audio spectrogram transformer (AST)~\cite{AST}, whose inputs are embeddings extracted from 16x16 patches on the 128-D mel-filter bank features with an overlap of 6. 
The BEATs checkpoint\footnote{\url{https://github.com/microsoft/unilm/tree/master/beats#beats}} is pre-trained and fine-tuned on AudioSet 2M.
The ECAPA-TDNN adopts a time delay neural network as the backbone and has a self-attentive pooling module to extract speaker information. The checkpoint\footnote{\url{https://huggingface.co/speechbrain/spkrec-ecapa-voxceleb}} trained on VoxCeleb1~\cite{Vox1} and VoxCeleb2~\cite{Vox2} was adopted. The model receives 80-D filterbank features as input and produces a single 192-D speaker embedding for a clip of audio. 
The performance of the teacher models are shown in Table~\ref{tab:singletask_baselines}.

\subsection{Student Model}

\noindent Zipformer~\cite{zipformer} is used as the architecture of the student encoder, and the medium configuration (Zipformer-M) is adopted. The encoder model consists of a stack of convolutional layers with ConvNeXT~\cite{convnext} followed by 6 encoder stacks with different downsampling ratios. The model receives 80-D filter bank features computed on 25~ms window with a stride of 10~ms and generates 512-D audio features at a frame-rate of 25~Hz. 
During pre-training, three branches for ASR, AT and SV tasks are added to the student encoder to compute the pre-training loss for each task. To address the frame-rate discrepancy between the Whisper teacher encoder and the student Zipformer encoder, successive Whisper embeddings are concatenated, forming a 2560-D feature vector at a frame-rate of 25~Hz. $\mathcal{M}_{\text{ASR}}$ is a linear projection layer from 512-D to 2560-D transforming the student embeddings to the same feature space of the Whisper teacher embeddings. 
$\mathcal{M}_{\text{AT}}$ consists of a classification layer followed by sigmoid activation, mapping the 512-D encoder features to 527 audio events. 
$\mathcal{M}_{\text{SV}}$ contains a attentive pooling module consisting of two 1D convolution layers, followed by a linear layer projecting the speaker embeddings to a 192-D vector.

During fine-tuning, the $\mathcal{M}_{\text{AT}}$ is not modified and the binary cross-entropy loss is adopted as the loss for AT training with labelled data. An additional classification layer is added after $\mathcal{M}_{\text{SV}}$ for speaker classification and cross-entropy loss is used for SV training. A predictor and joiner of the neural transducer are added onto the pre-trained student encoder for ASR. Note that the projection layer for learning Whisper embeddings is also retained during fine-tuning. Pruned-RNNT~\cite{prunedRNNT} loss, an efficient variant of the RNN-T loss function is used for ASR. 

\subsection{Dataset and Metrics}

\noindent The following datasets were used to pre-train and fine-tune the student audio encoder on ASR, AT and SV.

\textbf{ASR}: LibriSpeech (LS) is an English dataset containing 960 hours of book-readings from Project Gutenberg. The word error-rate (WER) is reported on the official test-clean and test-other testing sets.

\textbf{AT}: The AudioSet dataset is a large-scale audio classification dataset. It consists of over 2-million audio clips collected from YouTube. Each clip is annoted with one or more labels from a set of 527 labels. The whole training set is divided into two subsets, including a class-wise balanced set (AS-20K) and a class-wise unbalanced set (AS-2M). The availability of AudioSet is constantly changing as some videos from YouTube are removed. The AudioSet version from CED~\cite{CED} was adopted in this work, which contains 1,904,746 clips for training (21,155 of which are AS-20K) and 18,299 clips for evaluation. The mAP is used as the evaluation metric for both pre-training and fine-tuning.

\textbf{SV}: VoxCeleb1~\cite{Vox1} and VoxCeleb2~\cite{Vox2} were adopted for speaker verification. VoxCeleb1 contains around 350 hours of speech from 1251 speakers, and VoxCeleb2 contains over 2200 hours of speech from 6112 speaker. The combination of both VoxCeleb1 and VoxCeleb2 is referred to as Vox1,2 for convenience. The equal-error-rate (EER) on the official test set of VoxCeleb1 is reported. In addition, the LibriSpeech dataset is also used for SV. Score-normalisation is not used during evaluation.

During batching, random samples from multiple datasets were drawn with a weight proportional to the dataset sizes. This ensures a consistent distribution of training samples from each task in each batch and data from each task is exhausted at approximately the same time. One epoch of training terminates when any of the datasets is exhausted. 


\subsection{Training Details}

\noindent Training is terminated in the first stage of KD pre-training after iterating over LS-960 or LS-100 for 90 times. In multi-task pre-training and fine-tuning, the LibriSpeech dataset is repeated to match the amount of data from the other two tasks. The dataset configuration is shown in Table~\ref{tab:dataset}. When AS-2M is used during multi-task training, LS-960 is repeated 5 times. When performing ASR and SV multi-task training, LS-960 is repeated twice. During pre-training, the last ten checkpoints are averaged and used as the initialisation for fine-tuning. After fine-tuning, the results are reported also on the average of the last ten checkpoints.
\begin{table}[h]
    \centering
    \caption{The dataset distribution when performing multi-task training. KD pre-training and supervised fine-tuning shares the same distribution}
    \label{tab:dataset}
    \begin{tabular}{cc}
    \toprule
       Training Tasks & Dataset distribution  \\
    \midrule
       ASR + AT  &  LS-960  + AS-20K \\
       ASR + AT  &  LS-960 * 5 + AS-2M \\
       ASR + SV  &  LS-960 * 2 + Vox2 \\
       ASR + AT + SV  &  LS-960 * 5 + AS-2M + Vox1,2 \\
    \bottomrule
    \end{tabular}
\end{table}

During fine-tuning, the parameters in the pre-trained audio encoder (including subsampling front-end module and encoder layers of Zipformer) were frozen for the first 12,000 steps. The whole model is then updated for 15 epochs, with the encoder parameters having one fifth of the base learning rate of the model. ScaledAdam~\cite{zipformer} was used to optimise the model parameters during both pre-training and fine-tuning. A base learning rate of 0.045 was chosen and the Eden~\cite{zipformer} Scheduler is used to decay the learning rate as training progresses. 

Two 32GB NVIDIA-V100 GPUs were used in the small scale experiments, while the other experiments used four 32GB NVIDIA-V100 GPUs. Each batch has approximately 1000 seconds of audio. The whole training and fine-tuning process was implemented using icefall\footnote{https://github.com/k2-fsa/icefall} and Lhotse~\cite{lhotse} was used for data preparation.

%% file: 5.Results.tex
\section{Experimental Results}
\label{sec: results}

\subsection{Baseline Models}

\begin{table}[H]
    \centering
    \caption{Performance of teacher models and baseline single-task student models on 3 tasks. Student models trained from scratch using labelled data without KD. WER on test-clean and test-other, mAP on the AudioSet eval set and EER on VoxCeleb1 test set reported for ASR, AT and SV, respectively.}
    \label{tab:singletask_baselines}
    \begin{tabular}{cccc}
    \toprule
    Task & Model &  Dataset & Metrics\\
    \midrule
    
    \multirow{3}{*}{ASR} & Whisper-large-v3  & ASR & 4.44/6.3 \\
    & Zipformer  & LS-100 &  6.1/16.11\\
    & Zipformer  & LS-960 &  2.24/5.46 \\
    
    \midrule
    \multirow{3}{*}{AT} & BEATs & AS-2M & 48.6 \\
    & Zipformer  & AS-20K & 19.6  \\
    & Zipformer  & AS-2M  & 44.0   \\
    
    \midrule
    \multirow{3}{*}{SV} & ECAPA-TDNN  & Vox1,2 & 0.9 \\
    & Zipformer  & LS-100 &  45.4 \\
    & Zipformer  & LS-960 &  19.54 \\
    & Zipformer  & Vox1,2 & 2.40 \\
    \bottomrule
    \end{tabular}
    
\end{table}

The results of teacher models and baseline single-task student models are shown in Table~\ref{tab:singletask_baselines}. A clear performance gap between the AT and SV teachers and the Zipformer student model can be observed. Since Zipformer was initially designed for ASR, this performance difference suggests that different model architectures and different types of input features are crucial to the performance on AT and SV. The Whisper teacher model performs slightly worse than the baseline student Zipformer ASR model. This could be caused by the different text normalisations adopted by Whisper during training.

\begin{table}[H]
    \centering
    \caption{Results of multi-task student models trained from scratch with supervised data. For SV, the third layer output of encoder layer is used as input. AT module and ASR modules are appended at the end of audio encoder.}
    \label{tab:multitask_baselines}
    \begin{tabular}{cccccc}
    \toprule
    \multicolumn{3}{c}{Training Data} & \multicolumn{3}{c}{Metrics} \\
    \cmidrule(lr){1-3} \cmidrule(lr){4-6}
     ASR &  AT& SV   &  WER (\%)$\downarrow$ & mAP (\%)$\uparrow$ & EER (\%)$\downarrow$ \\
     \midrule
     LS-100   &  AS 20K  & -       & 5.68/15.21 & 25.2 & - \\
     LS-100   &  -  & LS-100        & 6.5/17.48 & - & 37.4 \\
     LS-960   &  AS 2M   & -       & 2.49/5.86  & 41.5 & - \\
     LS-960   &  -       & LS-960  & 2.72/6.78  &  -   & 12.56 \\
     LS-960   &  -       & Vox1,2    & 3.39/8.73 & - & 4.31 \\
     LS-960   &  AS 2M   & Vox1,2  & 2.71/6.63 & 35.6 & 5.10\\
     \bottomrule
     
    \end{tabular}
    
\end{table}

The results of the multi-task models trained from scratch with labelled data are shown in Table~\ref{tab:multitask_baselines}. The following observations can be made.
First, multi-task training of ASR and SV may improve the SV performance, but not vice versa. Multi-task training of ASR and SV on LS-100 and LS-960 result in EERs of 30.4 and 8.8, which are lower than the single-task SV models in Table~\ref{tab:singletask_baselines}. However, the WERs increased over 20\% relative, from 2.25/5.46 to 2.72/6.78 on LS-960, indicating that SV is negatively affecting ASR when trained with supervised data from scratch.
Second, multi-task training of ASR and AT leads to a performance improvement for AT if the training data for AT is limited. The multi-task ASR and AT model trained on a combination of LS-100 and AS-20K yielded a mAP of 25.2, whereas the single-task AT model only gave a mAP of 19.6. However, when sufficient AT data is used (AS-2M), the multi-task training of ASR and AT fails to improve the over the single-task AT baseline. When trained on LS-960 and AS-2M, the performance on both tasks are weaker compared to their single-task baselines.
Lastly, when all three tasks are trained together, the performance degradation on each task is bigger compared to the single-task or dual-task models, suggesting that multi-task training from scratch for the three tasks leads to poor task synergy and undermines single-task performance.

\subsection{Single-task KD pre-training}

\noindent The results of performing single-task KD training of the student model are shown in Table~\ref{tab:SingleKD}. Due to the pre-training loss selection for AT and SV, the student model is already capable of performing AT and SV. Therefore, mAP and EER are used as the evaluation metric for AT and SV. The L1 loss between the teacher Whisper and student embeddings computed on the dev-clean and dev-other set is used to measure the KD training performance for ASR. 
As shown in Table~\ref{tab:SingleKD}, the KD performance improves as the amount of training data grows. 
The student model trained with KD using the unlabelled AS-2M dataset outperforms the model trained with the same amount of labelled data, which is consistent with the findings in ~\cite{CED, AudioTaggingKD}. Since AudioSet is a multi-class classification task and many of the classes are correlated, the soft teacher label conveys richer information than the hard label. This also implies that AT models leveraging frame-level features are capable of achieving good results. 
For SV, the student model trained with KD obtained a lower EER than training from scratch with labelled data. We hypothesise that learning the speaker embedding directly by maximising the cosine similarity between the teacher and student speaker embeddings is an easier and more direct objective than classifying a speaker. 

\begin{table}[H]
    \centering
    \caption{Results of performing single-task KD on the student model on ASR, AT and SV. L1 loss on the Whisper embeddings, mAP (\%) and EER (\%) reported as metrics for ASR, AT and SV, respectively. }
    \label{tab:SingleKD}
    \begin{tabular}{cccc}
    \toprule
    \multicolumn{2}{c}{KD Training Config} & \multicolumn{2}{c}{Performance}\\
    \cmidrule(lr){1-2} \cmidrule(lr){3-4}
    Training Task & Dataset & Metrics & Results\\
    \midrule
    \multirow{2}{*}{ASR} & LS-100 &  \multirow{2}{*}{L1 error$\downarrow$} & 0.265    \\
     & LS-960 &  &  0.248    \\
     \midrule
    \multirow{2}{*}{AT} & AS-20K & \multirow{2}{*}{mAP(\%)$\uparrow$} & 26.4   \\ 
     & AS-2M &   & 47.8  \\
     \midrule
    \multirow{2}{*}{SV} & LS-960 &  \multirow{2}{*}{EER(\%)$\downarrow$} &  1.8 \\
     & Vox1,2 &  & 1.06 \\
    \bottomrule
    \end{tabular}
    
\end{table}

\subsection{Consistency of Data Augmentation}

\noindent The effect of applying different augmentations to the student model input data was investigated here and the results are shown in Table~\ref{tab:augmentation}. 
Different augmentations applied to the input of student model input when performing KD give rise to different effects on different tasks. Applying SpecAug~\cite{specaug} to the input features of the student model lead to weaker KD performance on ASR and SV, but slightly improves AT performance. 
Mixing input features with Musan~\cite{musan} for noise augmentation leads to better SV performance, but undermines the KD result on ASR and AT. 
Since the teacher targets were obtained using the un-augmented inputs, not applying augmentations to the student model input ensures consistency during KD training of the student model. Note that consistent data augmentation is possible in on-line KD, but the training efficiency is greatly affected due to the extra computation cost of teacher models. To minimise the KD performance reduction on all three tasks, no augmentation was performed to the student model input in the rest multi-teacher KD experiments.

\begin{table}[h]
    \centering
    \caption{Results of applying different data augmentations to student model input when performing single-task KD training.}
    \label{tab:augmentation}
    \begin{tabular}{ccccc}
        \toprule
         SpecAug & Musan & $\mathcal{L}^{\text{KD}}_{\text{ASR}}$ $\downarrow$ & mAP (\%) $\uparrow$ & EER (\%)$\downarrow$ \\
         \midrule
         \xmark & \xmark & \textbf{0.260}  & 29.0  & 9.44 \\
         \cmark & \xmark & 0.2738  & \textbf{29.9}  &  17.31 \\
         \xmark & \cmark & 0.262  & 28.0 & \textbf{9.10}  \\
         \cmark & \cmark & 0.2751  & 28.0  & 16.77 \\ 
         \bottomrule
    \end{tabular}
\end{table}

\subsection{Multi-task KD}

\subsubsection{ASR + AT}
\label{sec:ASR+AT}
Multi-task KD training experiments for ASR and AT were performed and results are shown in Table~\ref{tab:ASR+AT}. As reported in Whisper-AT, the last few layers of the Whisper model encodes the sound event information, suggesting that ASR and AT are not affecting each other. Since Whisper is selected as the ASR teacher, the embeddings of the last layer of the student encoder were used as the input to the AT module. As shown in Table~\ref{tab:ASR+AT}, performing multi-task KD on ASR and AT leads to only a minor degradation on each task.

\begin{table}[H]
    \centering
    \caption{Results of performing multi-task KD on ASR and AT.}
    \label{tab:ASR+AT}
    \begin{tabular}{cccc}
    \toprule
    \multicolumn{2}{c}{KD Training Data} & \multicolumn{2}{c}{Metrics} \\
    \cmidrule(lr){1-2} \cmidrule(lr){3-4}
     ASR  & AT & $\mathcal{L}^{\text{KD}}_{\text{ASR}}$ & mAP(\%) \\
     \midrule
        -      & AS-20K & - & 29.1 \\
        -      & AS-2M  & - & 47.8 \\
     LS-960    & AS-20K & 0.248 & 28.0 \\
     LS-960    & AS 2M  & 0.250 & 47.5     \\
     \bottomrule
    \end{tabular}
\end{table}

\subsubsection{ASR + SV}

\begin{table}[H]
    \centering
    \caption{Results of performing multi-task KD on ASR and SV, where SV module used $i_{\text{SV}}$-th layer output of encoder as input feature. Number of parameters in computation of SV also shown. Experiments carried out on LS-100.}
    \label{tab:sv_layers}
    \begin{tabular}{cccccc}
    \toprule
    \multirow{2}{*}{$i_{SV}$} & \multirow{2}{*}{\# SV Param} & \multicolumn{2}{c}{Training Data} & \multicolumn{2}{c}{Metrics} \\
    \cmidrule(lr){3-4} \cmidrule(lr){5-6}
       &  & ASR  &  SV & $\mathcal{L}^{KD}_{ASR}$ $\uparrow$ & EER (\%)$\downarrow$ \\
       \midrule
        - & -     & LS-100   &   -     & \textbf{0.260}  & - \\
       6  & 65.8M  &  -  &   LS-100      &    -    & 13.4 \\
       2  & 6.4M  & LS-100   &  LS-100  & \textbf{0.260}  & 38.0\\
       3  & 18.9M & LS-100   &  LS-100  & 0.262  & \textbf{9.44} \\
       4  & 49.2M & LS-100   &  LS-100  & 0.261  & 42.92    \\
       6  & 65.8M & LS-100   &  LS-100  & 0.261  & 45.44\\
       
    \bottomrule
    \end{tabular}
\end{table}
In contrast to Sec~\ref{sec:ASR+AT}, where using pure speech data for training AT is less informative, ASR and SV can share the same training set. Therefore, the speaker embeddings of LS-960 were also extracted so that they can be shared for KD training of ASR and SV. The Whisper embeddings are not extracted on Vox1 or Vox2. 
First, the relationship between ASR and SV was investigated and the SV module was put at different positions of the student encoder during multi-task KD. The L1 loss on the Whisper embeddings and EER on the Vox1 test set are reported in Table~\ref{tab:sv_layers}. 
Using different input features for the speaker module has a significant impact on the L1 loss and EER. Performing KD for ASR and SV together using the encoder features from the last layer results in significantly higher EER than the single task baseline, indicating that SV is strongly affected by ASR. This is expected as ASR should produce the same transcription given the same sentence spoken by different speakers, whereas SV aims to distinguish different speakers regardless of the content. 
Using the output of the intermediate blocks as input to the speaker module alleviates the issue: using the third block achieved the best trade-off between L1 loss and EER, suggesting that speaker information are preserved at intermediate layers. Hence, the remainder of this paper uses the third encoder block output as input features for speaker modules. 

Table~\ref{tab:asr+sv} shows the multi-task performance of ASR and SV on larger scale data. After placing the speaker module after the third block of the student encoder, multi-task KD of ASR and SV achieves comparable performance to the single-task KD models.

\begin{table}[H]
    \centering
    \caption{Results of performing multi-task KD on ASR and SV. $i_{SV}$ is fixed to 3.}
    \label{tab:asr+sv}
    \begin{tabular}{ccccc}
    \toprule
     \multicolumn{2}{c}{Training Data} & \multicolumn{2}{c}{Metrics} \\
    \cmidrule(lr){1-2} \cmidrule(lr){3-4}
    ASR Data  &  SV Data & $\mathcal{L}^{\text{KD}}_{\text{ASR}}$ $\uparrow$ & EER (\%) $\downarrow$ \\
    \midrule\
            LS-960   &  - & 0.248 & -  \\
           -     &  LS-960 & - &  1.80 \\
           LS-960   &  LS-960 & 0.249  & 1.80  \\
           LS-960   &  Vox1,2 & 0.2504 & 1.46\\
    \bottomrule 
    \end{tabular}
\end{table}

\subsubsection{ASR + AT + SV}
Here, the multi-task KD training for three tasks was carried out and results are reported in Table~\ref{tab:three tasks KD}. The mAP and EER consistently improve as the amount of unlabelled data for KD training for AT and SV increased. More particularly, combining LS-960 and Vox1,2 for SV resulted in the best SV performance, achieving EER of 1.13. The L1 error on the Whisper teacher embeddings only slightly increased as more data for SV and AT are used.

\begin{table}[H]
    \centering
    \caption{Results of performing multi-task KD on all three tasks. }
    \label{tab:three tasks KD}
    \adjustbox{max width=\linewidth}{
    \begin{tabular}{cccccc}
    \toprule
    \multicolumn{3}{c}{KD Training Data} & \multicolumn{3}{c}{Metrics} \\
    \cmidrule(lr){1-3} \cmidrule(lr){4-6}
     ASR  & AT & SV & $\mathcal{L}^{\text{KD}}_{\textbf{ASR}}$ $\downarrow$ & mAP(\%) $\uparrow$ & EER (\%) $\downarrow$ \\
     \midrule
     LS-960    & AS 20K & LS-960 & 0.250  & 28.3 &  1.79 \\
     LS-960    & AS 2M  & Vox1,2  & 0.251 & 46.8 & 1.48 \\
     LS-960    & AS 2M  & LS-960+Vox1,2  & 0.251 & 46.9 & 1.13 \\
     \bottomrule
    \end{tabular}    
    }
    
\end{table}

\subsection{Fine-tuning}

Although KD pre-training is performed using unlabelled data, the student model also possesses the capability of performing AT and SV due to the KD loss function for these two tasks after KD. However, the whole system still needs to be fine-tuned with ASR data. The main goal during the fine-tuning is to give the model good ASR performance while having minimal impact on AT and SV. Experiments were carried out on  with different initialisations. For easier reference, the model IDs of the KD pre-trained models are listed in Table~\ref{tab:alias}.

\begin{table}[H]
    \centering
    \caption{The IDs of the KD pre-trained models.}
    \label{tab:alias}
    \begin{tabular}{cccc}
    \toprule
      \multicolumn{3}{c}{Pre-training Data}  & \multirow{2}{*}{Model ID} \\
      \cmidrule(lr){1-3}
      ASR & AT & SV \\
      \midrule
      LS-960 & - & -   & \textbf{ST1} \\
      LS-960 & AS2M & Vox1,2   & \textbf{ST2} \\
      LS-960 & AS2M & Vox1,2+LS-960  & \textbf{ST3} \\
    \bottomrule
    \end{tabular}
\end{table}

\noindent  Since parameters involving SV only make up a small proportion of the whole encoder, one option would be freezing the parameters involving SV and therefore only the rest of the model for other tasks are fine-tuned. 
Also, it is shown in Table~\ref{tab:SingleKD} that using KD achieves better performance on SV and AT than supervised training, the KD loss can be still be adopted when fine-tuning the whole model on ASR tasks. 
The results of different fine-tuning strategies for SV are shown in Table~\ref{tab:finetune_mtl}. All models were fine-tuned with LS-960 for ASR, AS-2M for AT and Vox1,2 for SV (if SV parameters are not frozen).

\begin{table}[H]
    \centering
    \caption{Results of multi-task fine-tuning using different fine-tuning strategies. \cmark under ``Full FT'' means whole model is updated, otherwise speaker-related parameters are frozen. \cmark under ``KD loss'' means KD loss is used during fine-tuning. }
    \label{tab:finetune_mtl}
    \begin{tabular}{ccccccc}
    \toprule
     \multirow{2}{*}{Init} & \multirow{2}{*}{Full FT}  & \multicolumn{2}{c}{KD loss}  &  \multicolumn{3}{c}{Metrics} \\
     \cmidrule(lr){3-4} \cmidrule(lr){5-7}
      & & SV & AT & WER(\%) $\downarrow$ & mAP(\%) $\uparrow$ & EER (\%) $\downarrow$ \\
    \midrule
     \multirow{4}{*}{\textbf{ST2}} & \cmark & \xmark & \xmark & 2.3/5.56 & 41.3 &  3.03 \\
     & \cmark & \cmark & \cmark & 2.6/7.0 & 46.0 &  2.83 \\
     & \xmark & - & \xmark &  2.36/5.77  & 43.0 &  1.48 \\
     & \xmark & - & \cmark & 2.39/5.76 &  46.0 &  1.48 \\
     \midrule
     \textbf{ST3} & \xmark & - & \cmark & 2.35/5.63 & 45.9 & 1.13 \\
    \bottomrule
    \end{tabular}
    
\end{table}

If the whole model is fine-tuned without using KD as an auxiliary loss, the model already outperforms the multi-task baseline model trained from scratch (last row in Table~\ref{tab:multitask_baselines}) by a large margin on all three tasks, validating the effectiveness of multi-teacher KD pre-training. If the KD loss is adopted as an auxiliary loss during fine-tuning, the mAP and EER are improved at the cost of a large WER increase. Note that the mAP and EER are worse than for the initialisation model. This aligns with the observation (see Table~\ref{tab:multitask_baselines}) that multi-task learning leads to weaker performance on individual tasks. 
In particular, it is found that SV performance drops significantly compared to the initialisation model. To address this problem, another fine-tuning strategy is explored, where the SV-related parameters are frozen during the fine-tuning process to maintain the same SV performance as the initialisation model after fine-tuning. As shown in the final two rows in Table~\ref{tab:finetune_mtl}, ASR performance of the partially fine-tuned model is slightly worse than the fully fine-tuned model, but the overall performance on the three tasks is better. 
Our final system uses the best pre-trained model (\textbf{ST3}) as an initialisation, and is partially fine-tuned with the ASR loss and KD loss for AT. The performance of the final system is shown in the last row of Table~\ref{tab:finetune_mtl}. It achieved WERs of 2.35/5.63, mAP of 45.9 and EER of 1.13. Compared to the best-performing single-task teacher models, the multi-task student model yields competitive performance on all three tasks while requiring far fewer parameters and reduced computation.

\subsection{Effectiveness of pre-training}

As shown in Table~\ref{tab:finetune_mtl}, the proposed 2-stage multi-task multi-teacher KD framework successfully improves the performance on all tasks. An ablation experiment was conducted and experiments are shown in Table~\ref{tab:ablation}. 
Since KD on ASR and AT yields better performance than supervised training, one student model (second row) was trained with multi-task learning from scratch, where the losses for AT and SV are computed with KD. When comparing the two models in the first section, it can be confirmed that KD improves the performance on AT and SV at the cost of reducing ASR performance. 
The model fine-tuned on multi-task data from \textbf{ST1} (third row) outperforms the multi-task baseline (first row) on all evaluation metrics, demonstrating the effectiveness of pre-training with the Whisper teacher model. Comparing the final two rows, it can be found that the multi-teacher KD pre-training on AT and SV leads to a better performance on AT and SV after fine-tuning with labelled data.

\begin{table}[H]
    \centering
    \caption{Results of different configurations of multi-task training. All models used the same amount of supervised data: LS-960, AS-2M and Vox1,2 for training.}
    \label{tab:ablation}
    \begin{tabular}{lccc}
    \toprule
     \multirow{2}{*}{Model}   &  \multicolumn{3}{c}{Metrics}\\
     \cmidrule(lr){2-4}
        & WER(\%) $\downarrow$ & mAP(\%) $\uparrow$ & EER (\%) $\downarrow$ \\
    \midrule
    \multicolumn{3}{l}{\textbf{No pre-train}} \\
    Naive multi-task & 2.71/6.63 &  35.6 &  5.10 \\
    Multi-task, SV and AT w/ KD & 3.4/9.6 & 42.5 & 2.09 \\
    \midrule
    \multicolumn{3}{l}{\textbf{With pre-train}} \\
    From \textbf{ST1}, full FT w/o KD  & 2.24/5.46 & 38.0 & 3.23 \\
    From \textbf{ST2}, full FT w/o KD  & 2.3/5.56 & 41.3 & 3.03\\
    \bottomrule
    \end{tabular}
\end{table}

%% file: 6.Conclusions.tex
\section{Conclusions}
\label{sec: conclusions}

\noindent This paper presents a novel two-stage multi-task learning framework to build a general-purpose speech and audio encoder that jointly performs three fundamental tasks: ASR, AT and SV. In the first stage, multi-teacher KD is applied to align the feature space of a student model to three teacher models specialised for each task using unlabelled data. In the second stage, multi-task supervised fine-tuning is performed by initialising the model from the first stage. Experiments show that the proposed framework outperforms a multi-task model trained from scratch with supervised data by a large margin. The final multi-task model achieives WERs (\%) of 2.35/5.63 on the LibriSpeech test sets, mAP of 45.9\% on AudioSet, and EER (\%) of 1.13 on the VoxCeleb1 test set, which is close to the best-performing single-task models while requiring far fewer parameters and less computation.